\documentclass[prl,aps,twocolumn,showpacs,preprintnumbers,amsmath,amssymb]{revtex4}

\usepackage{graphicx}
\usepackage{dcolumn}
\usepackage{color}

\begin{document}

\def\EF{$E_\textrm{F}$}
\def\cred{\color{red}}
\def\cblue{\color{blue}}
\definecolor{dkgreen}{rgb}{0.31,0.49,0.16}
\def\cgreen{\color{dkgreen}}

\title{Study for material analogs of FeSb$_{2}$: material design for thermoelectric materials}

\author{Chang-Jong Kang$^1$}
\email[]{ck620@physics.rutgers.edu}
\author{Gabriel Kotliar$^{1,2}$}
\affiliation{
$^1$Department of Physics and Astronomy, Rutgers University,
Piscataway, New Jersey 08854, USA \\
$^2$Condensed Matter Physics and Materials Science Department,
Brookhaven National Laboratory, Upton, New York 11973, USA
}
\date{\today}

\begin{abstract}
Using the \emph{ab initio} evolutionary algorithm (implemented in USPEX)
and electronic structure calculations
we investigate the properties of a new thermoelectric material FeSbAs,
which is a material analog of the enigmatic thermoelectric FeSb$_{2}$.
We utilize the density functional theory and the Gutzwiller method
to check the energetics.
We find that FeSbAs can be made thermodynamically stable above $\sim$30 GPa.
We investigate the electronic structure and thermoelectric properties of FeSbAs
based on the density functional theory
and compare with those of FeSb$_{2}$.
Above 50 K, FeSbAs has higher Seebeck coefficients than FeSb$_{2}$.
Upon doping,
the figure of merit becomes larger for FeSbAs than for FeSb$_{2}$.
Another material analog FeSbP, was also investigated,
and found thermodynamically unstable even at very high pressure.
Regarding FeSb$_{2}$ as a member of a family of compounds (FeSb$_{2}$, FeSbAs, and FeSbP)
we elucidate what are the chemical handles that control the gaps in this series.
We also investigate solubility (As or P for Sb in FeSb$_{2}$) we found As to be more soluble.
Finally, we study a two-band model for thermoelectric properties
and find that the temperature dependent chemical potential
and the presence of the ionized impurities
are important to explain the extremum in the Seebeck coefficient
exhibited in experiments for FeSb$_{2}$.
\end{abstract}


\maketitle

\section{Introduction}

The search for thermoelectric materials having large Seebeck coefficients
has attracted lots of interest during the past several decades.
Since these materials have efficiency
for converting temperature differences in electric voltages,
it could be used to make refrigerators or power generators \cite{Rowe83}.
Correlated semiconductors and Kondo insulators containing
rare-earth or transition metal atoms have been regarded as
the possible candidates for good thermoelectric materials
due to a sharp singularity in the density of states
very near to the chemical potential \cite{Mahan96,Mahan98,Saso02,Grenzebach05}.

In order to get high efficiency or performance
in the thermoelectric materials for applications,
it is necessary to increase the (dimensionless) figure of merit as much as possible
\begin{equation}
ZT = \frac{\sigma S^2}{\kappa_{e}+\kappa_{l}}T,
\label{Seebeck}
\end{equation}
where $\sigma$ is the electrical conductivity,
$S$ is the Seebeck coefficient (also known as thermoelectric power),
$\kappa_{e}$ and $\kappa_{l}$ are the thermal conductivities
that are contributed from the electronic part and the lattice part, respectively.
Note that the numerator in Eq. (\ref{Seebeck}) is called as
the thermoelectric power factor ($PF = \sigma S^2$).
Therefore, the large thermoelectric power factor with the small thermal conductivities
gives the high figure of merit.

A correlated semiconductor FeSb$_{2}$ was reported
to have a gigantic Seebeck coefficient $S = -45$ mV/K at $\sim$12 K \cite{Bentien07,Sun10},
which results in the largest $PF$ ever found
($\sim$65 times larger than the $PF$ of the state-of-the-art thermoelectric
Bi$_{2}$Te$_{3}$-based material \cite{Mahan98}).
However, the lattice thermal conductivity $\kappa_{l}$ reaches a maximum
as large as $\sim$500 Wm$^{-1}$K$^{-1}$ at $\sim$15 K \cite{Bentien07,Sun10}.
It leads to a quite small $ZT$ value of 0.005 at the maximum of the $PF$.
Considering the fact that
any material with $ZT > 1$ is of great technological interest,
it is worth to try to reduce $\kappa_{l}$ significantly
without seriously affecting the $PF$.

The previous theoretical studies of FeSb$_{2}$ reported that
the gigantic Seebeck coefficient at low temperature
could not be described within both the density functional theory (DFT) level
and a local electronic picture \cite{Diakhate11,Tomczak10},
suggesting the substantial effect due to phonon drag \cite{Herring54,Takahashi16}
or the importance of vertex corrections \cite{Tomczak10},
the latter was attributed to the impurity band \cite{Battiato15}.
In this article, we have focused on the moderate temperature range (above $\sim$ 50 K)
where vertex corrections or phonon drag effect are not so important
and could be safely ignored.
Our aim in this work is to explore structural analogs of FeSb$_{2}$.
A natural question is, what will be the effect of substituting Sb by P or As?
We find that the As substitution is much more favorable.
At ambient pressure As is more soluble than P.
High pressure can stabilize FeSbAs,
but FeSbP is not stable even at very high pressure.
P substitution decreases the gap and the thermoelectric properties,
however, As substitution increases the gap and the Seebeck coefficient
and is a good target for synthesis. In the process of designing these materials we identify the
octahedral rotations that control the size of the band gap in the marcasite structure.

Modern theoretical methods of structure prediction  have been  very successful in  finding new  interesting materials experimentally.
Even though DFT has computational errors to determine the formation energy of a compound \cite{Hautier12,Kirklin15},
various corrections have been designed and implemented
in searchable repositories of DFT databases such as
Material Project \cite{Jain13}, OQMD \cite{Kirklin15}, and AFLOWlib \cite{Setyawan11}.
They give useful guideline to experiments for material synthesis and design.
Notable recent successes are the prediction of the 112 family of iron based superconductors
\cite{Shim09,Katayama13,Kang17}
and the prediction of superconductivity of hydrogen sulfide that has the highest critical temperature  under high pressure \cite{Duan14,Drozdov15}.
Other successes of theory guided material searches are the prediction  and synthesis of unreported missing half-Heusler compounds,
which are potential transparent conductors,
thermoelectric materials and topological semimetals \cite{Zhang12,Yan15,Gautier15}.
In addition,
new high-pressure phase materials
such as FeO$_{2}$ \cite{Hu16}, calcium carbides \cite{Li14}, and Na$_{2}$He \cite{Dong17}
were predicted by theory and confirmed to exist by experiment.
(For details of computational predictions based on DFT,
see the review \cite{Jain16}.)
This undoubtedly very partial list of accomplishments and references,
shows the speed at which theory is becoming
predictive and playing an important role in the search for new materials,
and here we employ this methodology to enlarge
the family of iron based marcasites and related structures.

\section{Method}

\begin{table}[b]
\caption{
Relaxed structural parameters of FeSb$_{2}$ (space group: $Pnnm$),
FeSbAs (space group: $Pmn2_1$), and FeSbP (space group: $Pmn2_1$).
The crystal axes for FeSbAs and FeSbP are
reoriented to have the space group of $Pn2_1m$
in order to compare with $Pnnm$ easily.
}
\begin{ruledtabular}
\begin{tabular}{c|c|c|c}
& FeSb$_{2}$ & FeSbAs & FeSbP \\
& $P$ = 0 & $P$ = 0 (40 GPa) & $P$ = 0\\
\hline
Space group & $Pnnm$ & Pn2$_{1}$m & Pn2$_{1}$m\\
$a ({\AA})$ & 5.761 & 5.555 (5.250) & 5.366\\
$b ({\AA})$ & 6.512 & 6.265 (5.884) & 6.097\\
$c ({\AA})$ & 3.297 & 3.051 (2.797) & 3.001\\
Wyckoff positions & & & \\
Fe & $2a$ & $2a$& $2a$ \\
& $x$ = 0.000 & $x$ = 0.774 (0.772) & $x$ = 0.792 \\
& $y$ = 0.000 & $y$ = 0.000 (0.000) & $y$ = 0.000 \\
Sb & $4g$ & $2a$ & $2a$ \\
&  $x$ = 0.198 & $x$ = 0.571 (0.574) & $x$ = 0.578 \\
& $y$ = 0.355 & $y$ = 0.365 (0.364) & $y$ = 0.375 \\
As (P) & $-$ & $2a$ & $2a$ \\
& & $x$ = 0.942 (0.937) & $x$ = 0.945 \\
& & $y$ = 0.651 (0.653) & $y$ = 0.656 \\
\end{tabular}
\label{struct_info}
\end{ruledtabular}
\end{table}

\begin{figure}[b]
\includegraphics[width=8.0 cm]{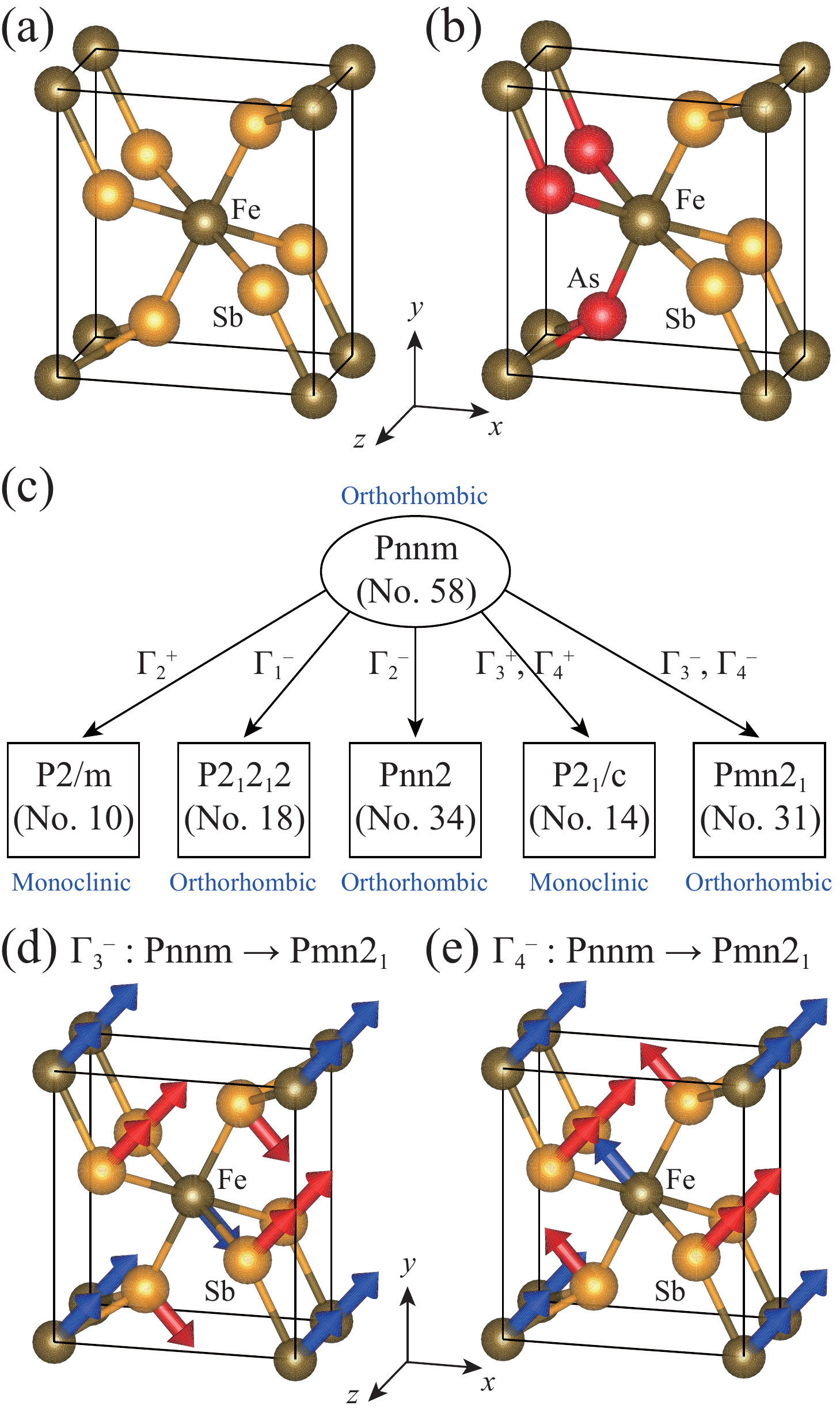}
\caption{(Color Online)
(a) Crystal structure of FeSb$_{2}$.
The space group is $Pnnm$ (No. 58).
(b) Crystal structure of FeSbAs.
The space group is $Pmn2_1$ (No. 31).
The crystal axes are reoriented for easily comparison with the crystal structure of FeSb$_{2}$.
The black lines represent the unit cell in both (a) and (b).
(c) Group table related to the Marcasite structure ($Pnnm$).
The structural phase transition between $Pnnm$ and $Pmn2_1$
is related to the (d) $\Gamma_{3}^{-}$ or (e) $\Gamma_{4}^{-}$ symmetries of the lattice distortions.
The arrows indicate atomic displacements in the reference structure of $Pnnm$.
}
\label{struct}
\end{figure}

To obtain the stable structural phase of FeSbAs,
we employ the \emph{ab initio} evolutionary algorithm \cite{Oganov06}
implemented in USPEX \cite{uspex}
combined with DFT
pseudopotential code VASP \cite{VASP1,VASP2}.
The initial structures are randomly generated
according to possible space groups.
In these calculations,
the structural optimization of all the newly generated structures
are carried out by VASP
with an energy cutoff of 500 eV and the exchange-correlation
functional of generalized gradient approximation (GGA) of
Perdew-Burke-Ernzerhof (PBE) \cite{PBE}
with the projector augmented wave (PAW) method \cite{PAW1,PAW2}.
After stable structural phases are obtained,
corresponding formation energies are calculated with
the Monkhorst-Pack sampling grid with a uniform density of 3,000 $k$-point per atom for the $k$-space integrations.

To check the energetics more precisely but relatively cheap,
we utilized the Gutzwiller method \cite{Lanata15,Gutz}
combined with the all-electron full-potential linearized augmented plane-wave (FLAPW) method implemented in WIEN2k \cite{Wien2k}.
We employ the general Slater-Condon parametrization of the on-site interaction
with $U$ = 5 and a Hund's coupling constant $J$ = 0.7 eV,
which were turned out to be reliable parameters in Fe compounds \cite{Tomczak12,Yin11}.

Both LDA (local-density approximation) and GGA(PBE) functionals
tend to underestimate band gaps of semiconductors.
To obtain the electronic structures with a reasonable band gap,
we utilized the modified Becke-John (mBJ) exchange potential \cite{Tran09},
which is rather accurate and computationally cheaper than $GW$ method.
We sampled the entire Brillouin zone (BZ) with 18$\times$16$\times$32 $k$ points.

The calculation of transport properties was performed using a denser mesh of
45$\times$40$\times$80 $k$ points of the BZ.
The semiclassical Boltzmann theory
as implemented in the transport code BOLTZTRAP \cite{Madsen06}
has been used to compute the electrical transport coefficients.

\begin{figure}[b]
\includegraphics[width=8.5 cm]{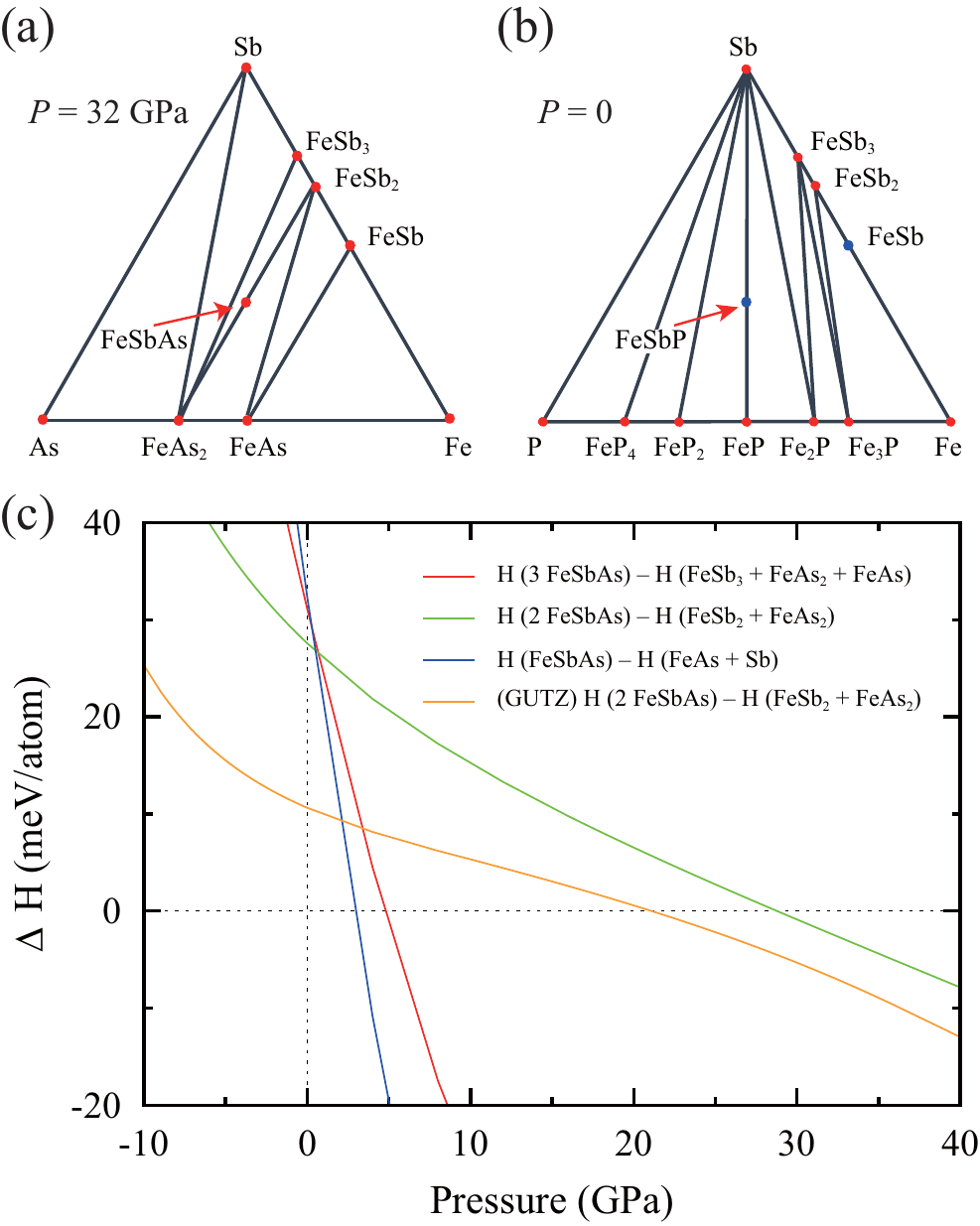}
\caption{(Color Online)
(a) Ternary phase diagram for FeSbAs under the pressure of 32 GPa.
FeSbAs is thermodynamically unstable at ambient pressure and
is only stable at high pressures ($P$ $\geq$ 30 GPa).
(b) Ternary phase diagram for FeSbP at ambient pressure.
Red and blue dots represent thermodynamically stable and unstable phase, respectively.
The energy above hull for FeSbP is 204.6 meV/atom.
Ternary phase diagrams are generated by pymatgen~\cite{Ong08,Ong10,pymatgen}.
(c) Enthalpy of formation for FeSbAs.
In order to obtain the enthalpy of formation as a function of pressure,
several lattice volumes including the optimized one were calculated,
and then the Murnaghan fitting was applied
to extract the enthalpy of formation at any pressure.
The relevant reaction for the stability of FeSbAs is
FeSb$_{2}$ + FeAs$_{2}$ $\rightarrow$ 2 FeSbAs.
We also checked the stability of this reaction
by using the GGA(PBE) + Gutzwiller (GUTZ) method.
The GUTZ method describes the smaller pressure to stabilize FeSbAs
compared to DFT GGA(PBE) method.
}
\label{ternary}
\end{figure}

\begin{table}[t]
\caption{
Enthalpy for each material
exhibited in Figs.~\ref{ternary}(a) and (b)
at $P$ = 0 or 32 GPa
calculated by VASP GGA(PBE) functional.
}
\begin{ruledtabular}
\begin{tabular}{c|c c}
& \multicolumn{2}{c}{Enthalpy $H = E + PV$ (eV/atom)} \\
& $P$ = 0 GPa & $P$ = 32 GPa \\
\hline
Fe & -8.236 & -6.100 \\
Sb & -4.138 & 1.185 \\
As & -4.672 & -0.784 \\
FeSb & -6.171 & -2.962 \\
FeSb$_{2}$ & -5.534 & -1.855 \\
FeSb$_{3}$ & -5.189 & -1.134 \\
FeAs & -6.681 & -4.181 \\
FeAs$_{2}$ & -6.123 & -3.313 \\
FeSbAs & -5.801 & -2.586 \\
\hline
P &  -5.404 & -1.242 \\
Fe$_{3}$P & -7.797 & -5.754 \\
Fe$_{2}$P & -7.767 & -5.718 \\
FeP & -7.425 & -5.294 \\
FeP$_{2}$ & -6.886 & -4.522 \\
FeP$_{4}$ & -6.345 & -3.664 \\
FeSbP & -6.125 & -3.132 \\
\end{tabular}
\label{enthalpy}
\end{ruledtabular}
\end{table}

\section{Computational results and discussion}

\subsection{Crystal structure and phase stability}
\begin{figure*}[t]
\includegraphics[width=18 cm]{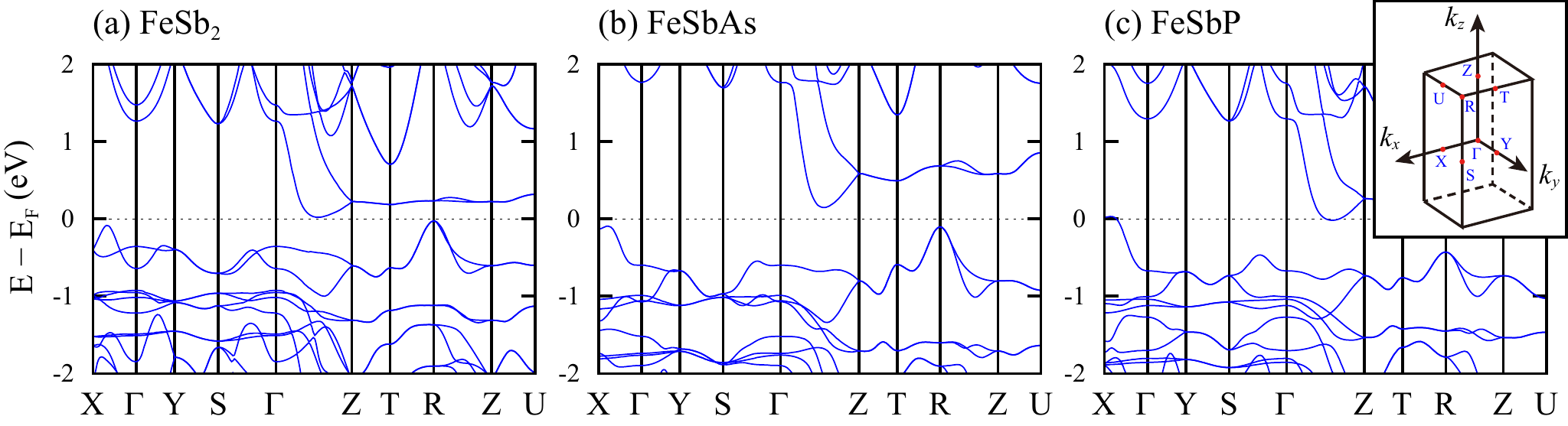}
\caption{(Color Online)
Band structures of (a) FeSb$_{2}$, (b) FeSbAs, and (c) FeSbP
calculated by the mBJ method.
Inset shows the bulk Brillouin zone.
The considered crystal structures are provided in TABLE. I.
For FeSbAs, the crystal structure at high pressure $P$ = 40 GPa was chosen.
}
\label{band}
\end{figure*}

The crystal structure of FeSb$_{2}$ is shown in Fig. \ref{struct}(a),
which has the orthorhombic marcasite structure (space group: $Pnnm$)
where a Fe ion is surrounded octahedrally by six Sb anions \cite{Rosenqvist53}.
The Fe octahedron has corner sharing with the neighboring Fe octahedron in an $x$-$y$ plane,
however it has edge sharing along the $z$ direction.
Therefore, it is expected to have larger band dispersion along the $z$ direction
than $x$ or $y$ directions (see Fig. \ref{band}).
The crystal structure of FeSbAs at ambient or under pressures founded by USPEX
has a space group $Pmn2_1$,
and it can be interpreted as the structural phase transition
from the space group $Pnnm$ with the $\Gamma_{3}^{-}$ or $\Gamma_{4}^{-}$ symmetries of the lattice distortions as shown in Fig. \ref{struct}(c).
These $\Gamma_{3}^{-}$ or $\Gamma_{4}^{-}$ lattice distortions are confined within the $x$-$y$ plane
and lead to break the inversion symmetry
(the replacement of three Sb with three As atoms
breaks the inversion symmetry as well).
The structural information on FeSbAs at ambient or under pressure could be found in
TABLE \ref{struct_info} along with FeSb$_{2}$ for comparison.

FeSbAs at ambient pressure is thermodynamically unstable
and decomposed into FeAs and Sb compounds.
The energy above hull is 32.2 meV/atom.
We have checked the phase stability of FeSbAs under pressures
dealing with the enthalpy $H = E + PV$,
where $E$ is the total electronic energy, $P$ is the external pressure,
and $V$ is the crystal volume.
Three relevant reactions for FeSbAs under pressures are shown in Fig. \ref{ternary}(c),
where the enthalpy of formation $\Delta H$ is plotted as a function of pressure.
The positive or negative enthalpy of formation indicates
that FeSbAs is thermodynamically unstable or stable, respectively.
Two reactions such as FeSb$_{3}$ + FeAs$_{2}$ + FeAs $\rightarrow$ 3 FeSbAs and
FeAs + Sb $\rightarrow$ FeSbAs are stable above $\sim$8 GPa,
however FeSb$_{2}$ + FeAs$_{2}$ $\rightarrow$ 2 FeSbAs is only stable at high pressure
($P \geq$ 30 GPa).
Therefore, the reaction process of FeSb$_{2}$ + FeAs$_{2}$ $\rightarrow$ 2 FeSbAs
is an indicator of the phase stability of FeSbAs
(the detail enthalpy for each material is listed in TABLE.~\ref{enthalpy}).
We also checked the phase stability of FeSbAs with the GGA(PBE) + Gutzwiller (GUTZ),
which is the advanced but relatively cheap method for doing energetics.
The GGA + GUTZ method describes the smaller pressure to stabilize FeSbAs compared to DFT GGA(PBE) method.
The LDA + GUTZ method is also tested because LDA + GUTZ gives better energetics than GGA + GUTZ.
With LDA + GUTZ, it gives the stable phase of FeSbAs above 30 GPa, which is quite close to the DFT GGA(PBE) result (not shown).
Hence, it is safe to mention that FeSbAs is only stable above $P \sim$ 30 GPa,
and FeSbAs is decomposed into FeSb$_{2}$ and FeAs$_{2}$ compounds below the pressure.

Even though FeSbAs is thermodynamically stable
only at high pressure (above $\sim$30 GPa),
we checked the phonon dispersion of FeSbAs at ambient pressure
and found that it is mechanically stable (no imaginary phonon softening).
It is the similar situation in diamond:
diamond is less thermodynamically stable than graphite,
however is mechanically stable at ambient pressure.
Once FeSbAs is synthesized at high pressures,
it could be released into ambient pressure
without decomposing into other substances
(assuming that the conversion rate from FeSbAs to other substances
is negligible at standard conditions).

Another material analog FeSbP is also tested.
First, USPEX was performed to search the crystal structure for FeSbP.
The space group Pmn2$_{1}$ (No. 31) was obtained, which is same as one for FeSbAs.
FeSbP at ambient pressure is thermodynamically unstable
and decomposed into FeP and Sb compounds (Fig.~\ref{ternary}(b)).
The energy above hull is 204.6 meV/atom,
which is much larger than that for FeSbAs at ambient pressure.
It indicates that FeSbP is more thermodynamically unstable than FeSbAs.
We have checked the phase stability of FeSbP under pressures
and found that FeSbP is unstable
over the whole pressure range (even at high pressure).

Hereafter, we will discuss the electronic structure
and thermoelectric properties for FeSbAs
at high pressure $P$ = 40 GPa unless otherwise noted.
For FeSbP, there is no any pressure to stabilize it,
but the ambient pressure phase is chosen to study the electronic structure
and thermoelectric properties in order to compare with other material analogs.

\subsection{Electronic structure}

Figure \ref{band} shows band structures of FeSb$_{2}$, FeSbAs, and FeSbP
with the structural parameters exhibited in TABLE \ref{struct_info}.
All the materials FeSb$_{2}$, FeSbAs, and FeSbP show the metallic phase
based on the GGA(PBE) functional.
The standard DFT describes severe underestimation of a bulk band gap,
so that DFT + $U$, hybrid functional, or $GW$ are required to obtain the proper bulk band gap.

First, we checked the GGA + $U$ (the on-site Coulomb repulsion parameter) method
with an effective $U_{eff} = U - J$ parameter setting $J = 0$.
For $U_{eff}$ = 4 eV, the band gap of FeSb$_{2}$ is 166 meV,
which is almost 5 times larger than the experimental band gap of $\sim$30 meV
\cite{Bentien07,Perucchi06,Petrovic03}.
While FeSb$_{2}$ is paramagnetic in experiments,
the GGA + $U$ method describes that
the ferromagnetic solution has the lower energy
than the nonmagnetic solution~\cite{Lukoyanov06}.
Therefore, the GGA + $U$ method is not suitable
to study the electronic structure of FeSb$_{2}$.

\begin{figure}[t]
\includegraphics[width=8.5 cm]{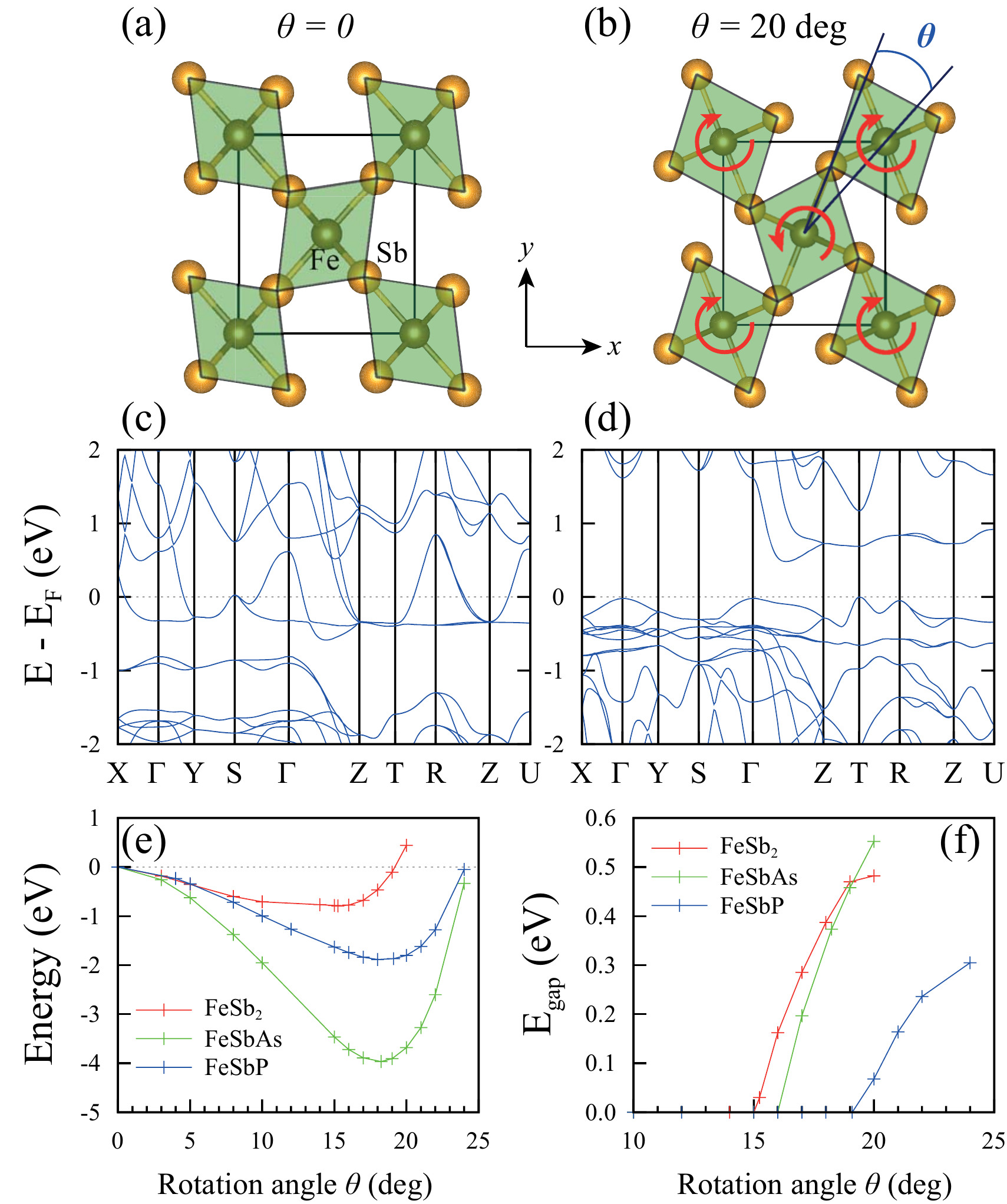}
\caption{(Color Online)
Hypothetical crystal structure of FeSb$_{2}$
with an octahedron rotation angle (a) $\theta$ = 0
and (b) $\theta$ = 20 deg
with keeping the space group $Pnnm$ (No. 58).
Fe (Sb) atoms are located in the center (corner)
of the green-colored octahedrons.
The corresponding electronic band structures
of (a) and (b) calculated by the mBJ method are
shown in (c) and (d), respectively.
The band gap in (d) is indirect and
its magnitude is 0.48 eV.
(e) Energy vs. rotation angle $\theta$ for FeSb$_{2}$, FeSbAs, and FeSbP
calculated by the mBJ method.
Energy for $\theta$ = 0 for each material is set to be zero as reference energy.
Equilibrium rotation angles for FeSb$_{2}$, FeSbAs, and FeSbP are
15.2, 18.2, and 19.1 degrees, respectively.
(f) Band gap vs. rotation angle $\theta$ for FeSb$_{2}$, FeSbAs, and FeSbP
calculated by the mBJ method.
The band gap could be controlled by the octahedron rotation.
}
\label{rotation}
\end{figure}

The mBJ method was thoroughly tested for many semiconductors and insulators
to obtain band gaps close to the experimental ones \cite{Tran09}.
The band gap of FeSb$_{2}$ described in mBJ is 19 meV,
which is very close to the experimental data of $\sim$30 meV.
Thus, in view of a much reduced computational time compared to the $GW$ \cite{Tomczak10},
mBJ is indeed an efficient method to get a reasonable band gap
and proper thermoelectric properties at the moderate temperature range above 50 K.
(At low temperature,
the vertex correction is inevitable to get the reasonable thermoelectric power.
Here, we have focused on the moderate temperature range above 50 K
where the vertex correction or phonon-phonon interaction are not so important.)
For the case of FeSbAs, mBJ gives the band gap of 146 meV,
which is quite large compared to the FeSb$_{2}$ case.
On the other hand, FeSbP is a compensated semimetal within the mBJ scheme.
The valence band along $\Gamma$-X and the conduction band along $\Gamma$-Z
cross the Fermi level, which results in giving hole and electron pockets, respectively.

The size of the band gap is closely related to a rotation angle $\theta$ (Fig.~\ref{rotation}(b))
of the Fe octahedron around the $z$ axis.
At zero rotation angle,
the band gap is closed as shown in Fig.~\ref{rotation}(c)
due to large hopping integrals between Fe and Sb atoms.
The band gap begins to open at a certain amount of the rotation angle
(so-called $\theta_{0}$),
and it increases further to get the maximum and then decreases
as the rotation angle increases.
The rotation angle $\theta_{0}$ could differ depending on the
anion size and the crystal unit cell volume.
The rotation angles in the equilibrium structures for FeSb$_{2}$, FeSbAs, and FeSbP
are 15.2, 18.2, and 19.1 degrees, respectively (Fig.~\ref{rotation}(e)).
If the rotation angle is increased further to be 20 deg in FeSb$_{2}$,
the band gap is also increased further and its magnitude is 0.48 eV (Figs.~\ref{rotation}(d) and (f)).
For FeSbP, the band gap starts to open at $\theta_{0} \approx 20$ deg,
which is larger than $\theta_{0}$ for other two materials
as shown in Fig.~\ref{rotation}(f).

Since a larger bulk gap material usually gives a larger high temperature thermoelectric power~\cite{Tomczak10},
it is expected that the thermoelectric power for FeSbAs is the largest,
FeSb$_{2}$ is the second, and FeSbP is the smallest among them.

\begin{table}[t]
\caption{
Physical parameters for FeSb$_{2}$, FeSbAs, and FeSbP.
Effective masses of valence ($m_{V}^*$) and conduction ($m_{C}^*$) bands
are calculated from the relation
$(m^*)_{ij}^{-1} = \frac{1}{\hbar^2}\frac{\partial^2 E(k)}{\partial k_i k_j}$
in the unit of the rest mass of an electron.
Average effective masses $m_{V}^*$ and $m_{C}^*$ are obtained via the harmonic mean.
The Debye temperatures $\theta_{D}$ are obtained from
the first-principle phonon calculations \cite{phonon}.
The sound velocities $v_{s}$ are calculated from a Debye model.
A phonon mean free path $l_{p}$ is assumed to be proportional to
an average lattice constant, that is, $l_{p} \propto \sqrt[3]{V}$,
where $V$ is the unit cell volume.
Given $l_{p}$ = 350 $\mu$m for FeSb$_{2}$ \cite{Bentien07},
$l_{p}$ for FeSbAs and FeSbP could be obtained.
}
\begin{ruledtabular}
\begin{tabular}{c|c|c|c}
& FeSb$_{2}$ & FeSbAs & FeSbP \\
\hline
($m_{V}^*$)$_{xx}$ & -0.21 & -0.40 & -0.79 \\
($m_{V}^*$)$_{yy}$ & -0.34 & -0.46 & -0.51 \\
($m_{V}^*$)$_{zz}$ & -0.43 & -0.44 & -8.97 \\
$m_{V}^*$          & -0.30 & -0.43 & -0.90 \\
\hline
($m_{C}^*$)$_{xx}$ & 1.43 & 0.82 & 1.19 \\
($m_{C}^*$)$_{yy}$ & 3.64 & 2.44 & 4.85 \\
($m_{C}^*$)$_{zz}$ & 1.39 & 0.96 & 1.86 \\
$m_{C}^*$          & 1.77 & 1.12 & 1.90 \\
\hline
$\theta_{D}$ (K) & 286 & 476 & 420 \\
$v_{s}$ (m/s)    & 2634 & 3890 & 3581 \\
$l_{p}$ ($\mu$m) & 350 & 311 & 324 \\
\end{tabular}
\label{effM}
\end{ruledtabular}
\end{table}

\subsection{Thermoelectric properties}

\begin{figure}[t]
\includegraphics[width=8.5 cm]{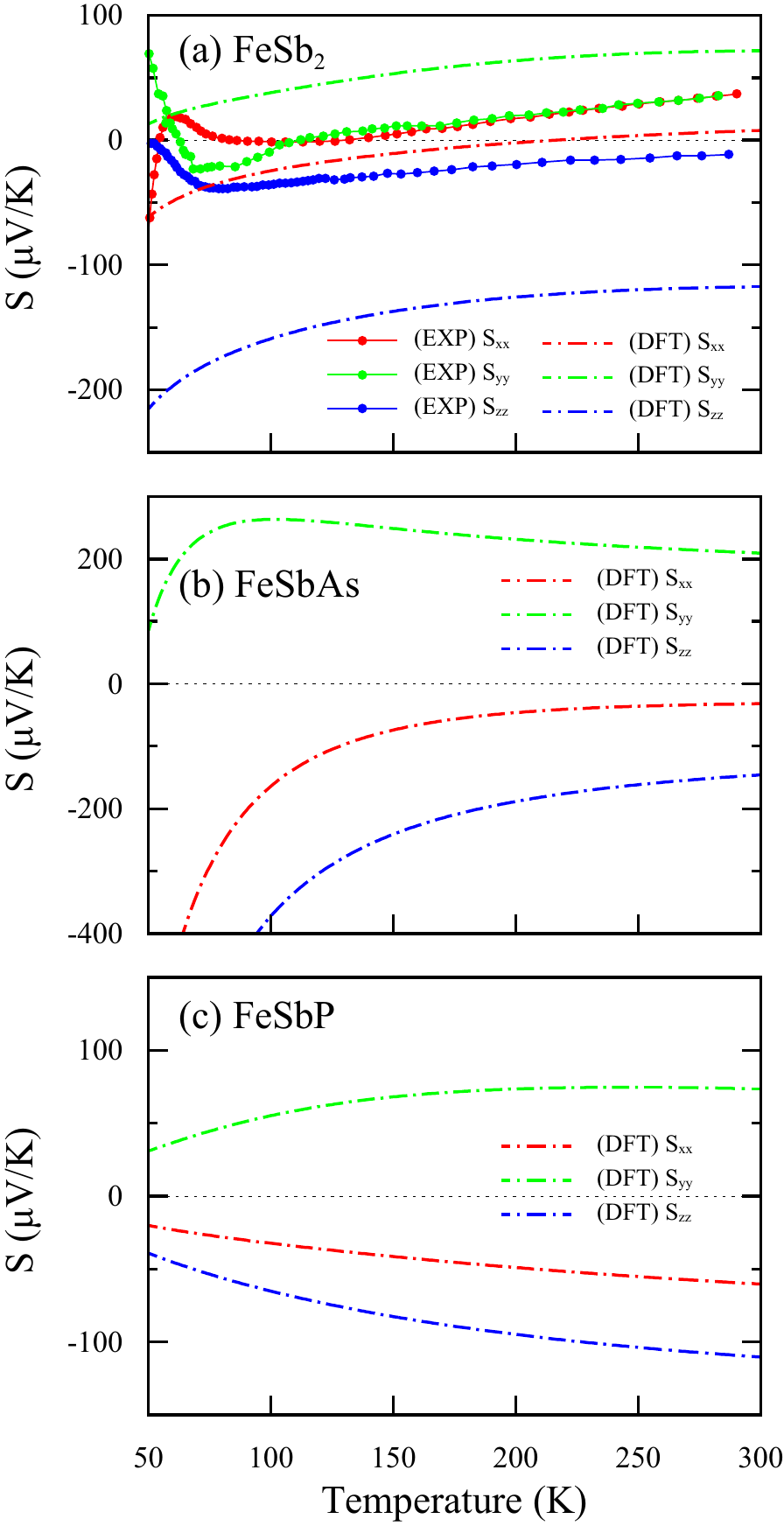}
\caption{(Color Online)
(a) Thermoelectric power of FeSb$_{2}$ as a function of temperature.
Experimental data from Ref. \cite{Bentien07} are shown for comparison with
DFT (mBJ potential method) results.
Thermoelectric powers of (b) FeSbAs and (c) FeSbP
calculated by DFT (mBJ method)
as a function of temperature.
The temperature range from 0 to 50 K is the unreliable region
within the current theoretical approach (see main text for details),
so that the moderate temperature region above 50 K is focused in this study.
}
\label{thermopower}
\end{figure}

The theoretically calculated thermoelectric power for FeSb$_{2}$ as a function of temperature is shown in Fig. \ref{thermopower}(a) with the experimental data for comparison.
We set the chemical potential to be the middle of the band gap.
At low temperature below 50 K, the theoretically calculated thermoelectric power could not describe the gigantic thermoelectric power observed in experiments.
This suggests that vertex corrections and nonlocal correlation effects that we neglect are important. The presence of a substantial phonon-drag effect could also give the huge inconsistency between experiments and the current theory.

Above $\sim$80 K, the calculated thermoelectric powers
are well matched with the experiment
in the sense that the thermoelectric powers
described by both theory and experiment have the same order of magnitude
(except for $zz$ component)
and have the same increasing tendency, that it, the same positive slope
in the thermoelectric power versus temperature curve.
However, there is a discrepancy in the thermoelectric power along $zz$ direction ($S_{zz}$)
between mBJ and the experiment:
mBJ describes $S_{zz}$ several times (from $\sim$4 to $\sim$8 times) larger
than the experiment at the temperature range between 100 and 300 K.
This discrepancy might come from the fixed chemical potential
over the temperature range in the calculation.
In addition, the several types of impurities such as electron donor or hole acceptor impurities
could be also important to give better consistency between theory and experiment.
We will discuss the temperature behavior of the thermoelectric power
with the chemical potential varied with temperature
and the effect of the presence of impurities
in Appendix.

\begin{figure}[t]
\includegraphics[width=8.5 cm]{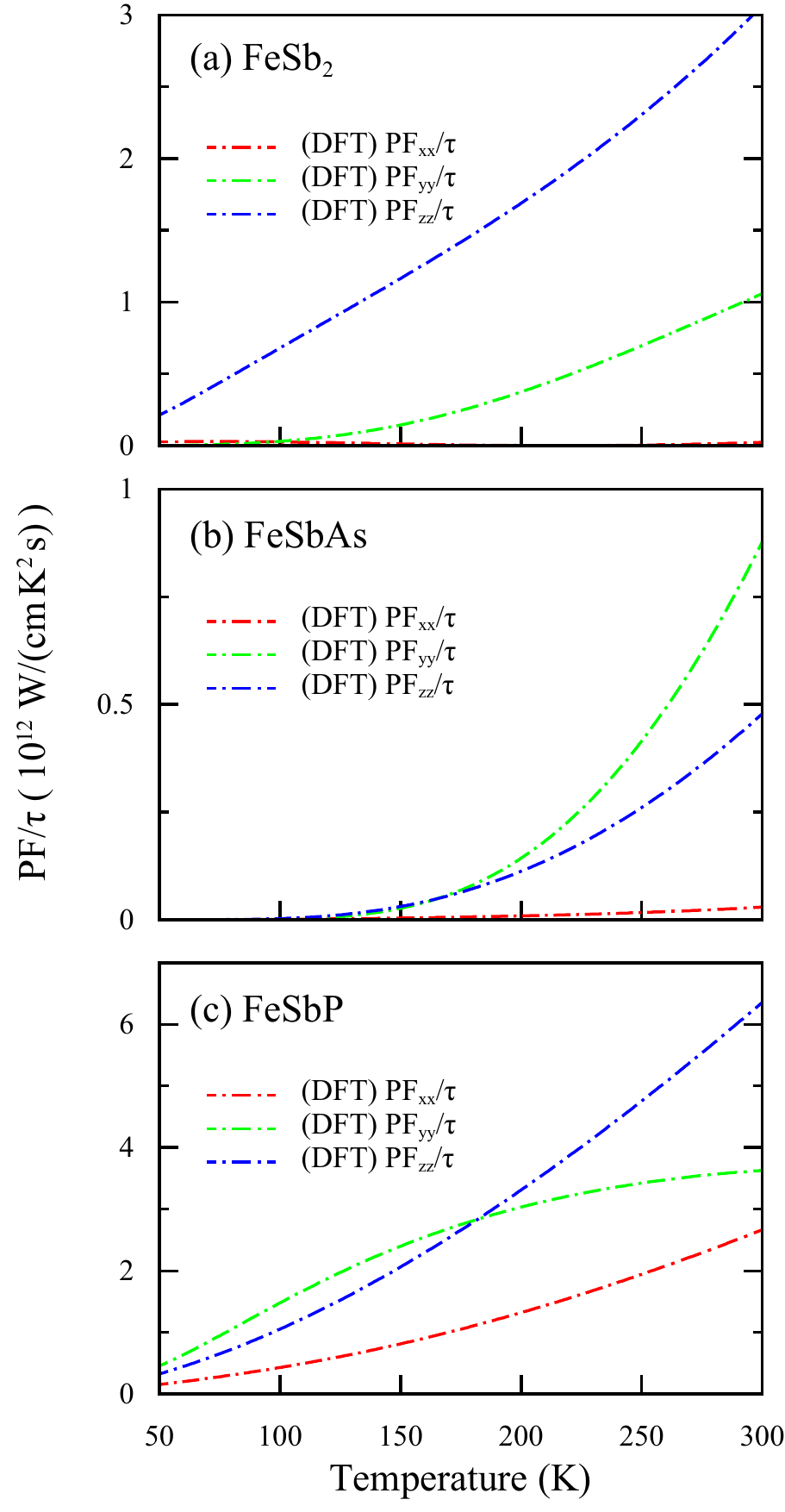}
\caption{(Color Online)
Power factors divided by the relaxation time (PF/$\tau$)
(calculated by the mBJ scheme)
for (a) FeSb$_{2}$, (b) FeSbAs, and (c) FeSbP
as a function of temperature.
}
\label{PF}
\end{figure}

\begin{figure}[t]
\includegraphics[width=8.5 cm]{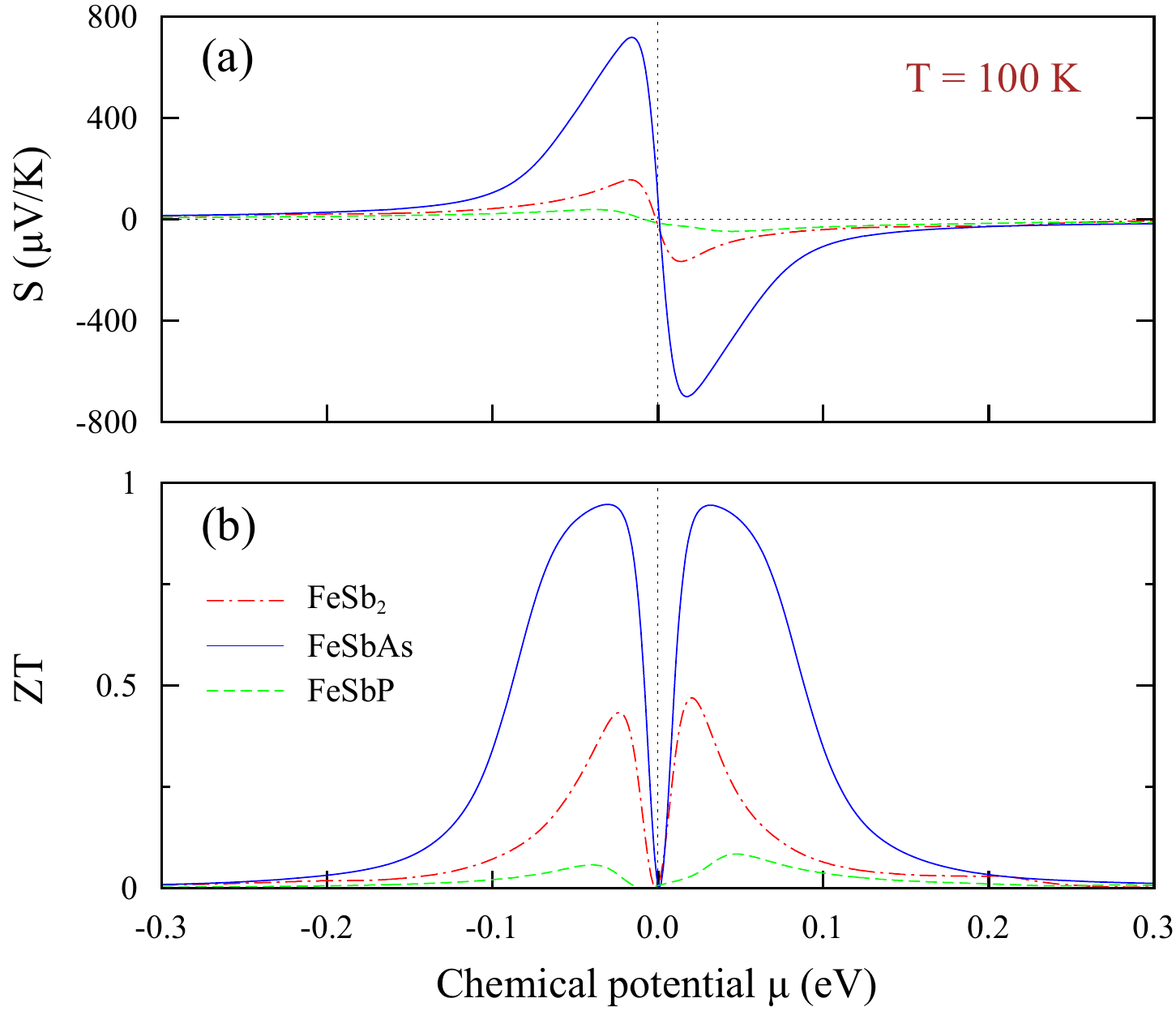}
\caption{(Color Online)
Comparison in thermoelectric properties
(calculated by the mBJ scheme)
among FeSb$_{2}$, FeSbAs, and FeSbP materials at $T$ = 100 K.
(a) Seebeck coefficients and (d) figure of merits $ZT$
(without considering the lattice thermal conductivities)
are shown as a function of the chemical potential.
These two thermoelectric properties
are averaged over the three $x$, $y$, and $z$ directions.
}
\label{thermopower_mu}
\end{figure}

The thermoelectric power for FeSbAs is also calculated and shown in Fig.~\ref{thermopower}(b).
The thermoelectric power for FeSbAs is much enhanced compared to FeSb$_{2}$:
$\sim$6.7, $\sim$6.9, and $\sim$2.3 times enhanced in a magnitude along $xx$, $yy$, and $zz$ directions, respectively, at $T$ = 100 K.
The Seebeck coefficients could be enhanced by the shift of the chemical potential,
supposing that electron or hole doping does not alter the electronic structure heavily.
The maximum value of the Seebeck coefficient
could be achieved with 0.01 $\sim$ 0.02 eV shift in the chemical potential.
Then the Seebeck coefficient is about 4 times larger
for FeSbAs than for FeSb$_{2}$
(see Fig.~\ref{thermopower_mu}(a)).
Hence, the newly proposed material FeSbAs has larger thermoelectric powers
than FeSb$_{2}$ above 50 K.

Power factors ($PF$s) divided by the relaxation time ($\tau$)
calculated within mBJ for FeSb$_{2}$
are larger than those for FeSbAs above 50 K (Fig.~\ref{PF}).
For example, at $T$ = 300 K,
they are 0.022 (0.03), 1.05 (0.88), and 3.05 (0.48)
$\times 10^{12}$ W/(cm K$^{2}$ s)
for $xx$, $yy$, and $zz$ directions, respectively, for FeSb$_{2}$ (FeSbAs).
Even though Seebeck coefficients for FeSbAs are larger than for FeSb$_{2}$ above 50 K,
the electrical conductivity ($\sigma$) for FeSbAs is smaller than for FeSb$_{2}$
due to the larger band gap,
so that the resulting $PF/\tau$ (=$\sigma S^2/\tau$) for FeSbAs
is smaller than for FeSb$_{2}$.
However, if the chemical potential could be shifted through the doping
without altering the electronic structure heavily,
$PF/\tau$ for both FeSbAs and FeSb$_{2}$ are quite comparable
at around 0.1 or around -0.1 eV shift of the chemical potential.

Since FeSbP has the metallic phase within the mBJ method,
the linear temperature dependent thermoelectric power ($S \sim T$) is demonstrated
at low temperature as shown in Fig.~\ref{thermopower}(c).
The thermoelectric powers for FeSbP along $xx$, $yy$, and $zz$ directions
are confined below $\sim$100 $\mu V/K$ in magnitude
at the temperature range between 50 and 300 K.
They are much reduced compared to those for FeSbAs
but are similar in size with those for FeSb$_{2}$ except for the $zz$ component.
FeSb$_{2}$ has the larger $zz$ component than FeSbP
and it leads to the larger average thermoelectric power for FeSb$_{2}$ than for FeSbP
as shown in Fig.~\ref{thermopower_mu}(a).

Even though thermoelectric powers for FeSbP are smaller than those for FeSbAs,
$PF/\tau$ for FeSbP is much larger than that for FeSbAs
($\sim$13 times larger at $T$ = 300 K for the $zz$ component)
due to the larger $\sigma$ of metallic FeSbP
than that of insulating FeSbAs.
However, $PF/\tau$ for both FeSbAs and FeSbP are quite comparable
for the chemical potential shift larger than +0.1 eV.

Figure~\ref{thermopower_mu}(b) shows the figure of merit $ZT$
(here, the only electronic contribution to the thermal conductivity is considered)
as a function of the chemical potential
for three different materials FeSb$_{2}$, FeSbAs, and FeSbP.
The figure of merit for FeSbAs is almost 2 times enhanced
compared to that for FeSb$_{2}$.
Considering that $PF/\tau$ for FeSbAs is smaller than for FeSb$_{2}$,
this enhancement is due to the smaller electronic thermal conductivity
($\kappa_{e}$) in FeSbAs than in FeSb$_{2}$.
(Since FeSbAs has a larger band gap than FeSb$_{2}$,
it makes both $\sigma$ and $\kappa_{e}$ of FeSbAs
smaller than those of FeSb$_{2}$.)
For FeSbP, both $\sigma$ and $\kappa_{e}$
are large due to the metallic phase.
It leads to a quite small figure of merit for FeSbP.

The electronic thermal conductivity could be calculated
from the electrical conductivity by using the Wiedemann-Franz law
with the Lorenz number $L_{0} = 2.44 \times 10^{-8}$ W$\Omega$K$^{-2}$.
From electrical conductivities of FeSb$_{2}$
measured at 100 K \cite{Bentien07},
electronic thermal conductivities of FeSb$_{2}$ at 100 K
along $xx$, $yy$, and $zz$ directions
are 0.168, 0.127, and 0.110 (0.262 for a different sample)
Wm$^{-1}$K$^{-1}$, respectively.
Compared to the lattice thermal conductivity ($\kappa_{l}$) of FeSb$_{2}$ measured at 100 K,
which is about 20 $\sim$ 30 Wm$^{-1}$K$^{-1}$ \cite{Bentien07,Sun10},
$\kappa_{e}$ is much smaller than $\kappa_{l}$.
Taking into account $\kappa_{l}$,
$ZT$ is much reduced to have the order of magnitude of $10^{-4} \sim 10^{-5}$.
Therefore, reducing $\kappa_{l}$
is an ultimate goal for increasing $ZT$.
Sun \emph{et al.} reported that
a slight substitution of As in FeSb$_{2}$, FeSb$_{2-x}$As$_{x}$ ($x$ = 0.03),
reduces the thermal conductivity
much more up to by a factor of 5 \cite{Sun10}.
The reduction is due to the effect of substitutional disorder.

\begin{figure}[t]
\includegraphics[width=8.5 cm]{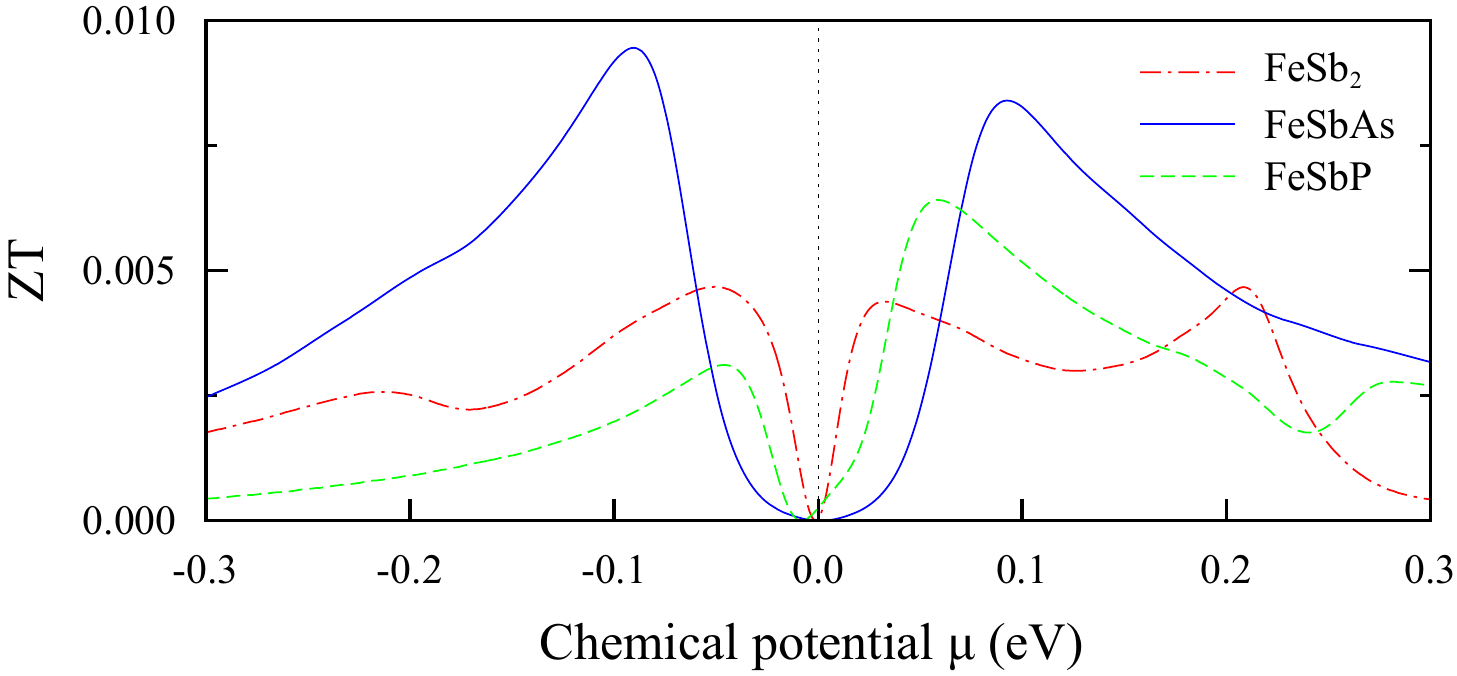}
\caption{(Color Online)
Figure of merits $ZT$ at $T$ = 100 K
(with considering the lattice thermal conductivities)
are shown as a function of the chemical potential.
}
\label{ZT}
\end{figure}

We can estimate $\kappa_{l}$
from the kinetic formula $\kappa_{l} = 1/3~C(T) \cdot v_s \cdot l_{p}$,
where $C(T)$, $v_s$, and $l_{p}$ are the lattice specific heat, sound velocity,
and phonon mean free path, respectively.
The lattice specific heat $C(T)$ and sound velocity $v_s$ could be calculated
from a Debye model with the Debye temperature listed in TABLE.~\ref{effM}.
With the phonon mean free path provided in TABLE.~\ref{effM},
the estimated lattice thermal conductivities
of FeSb$_{2}$, FeSbAs, and FeSbP at 5 K are
257, 105, and 129 Wm$^{-1}$K$^{-1}$, respectively.
Scaling the measured $\kappa_{l}$ of FeSb$_{2}$ at 100 K
by the ratio of the estimated lattice thermal conductivities at 5 K,
we can roughly estimate the lattice thermal conductivities of FeSbAs and FeSbP at 100 K
to be $\sim$10 and $\sim$12 Wm$^{-1}$K$^{-1}$, respectively.
Using these values,
$ZT$ at 100 K \cite{tau} is plotted in Fig.~\ref{ZT}.
Again, the newly proposed material FeSbAs has a higher $ZT$
than FeSb$_{2}$ with a chemical potential shift,
which is accomplished by electron or hole doping.

We also checked the electronic structure
and thermoelectric properties of FeSbAs at ambient pressure.
The insulating phase cannot be obtained in the GGA(PBE) functional.
The mBJ method describes the insulating phase
with the band gap of 297 meV.
This value is almost two times larger than that at high pressure $P$ = 40 GPa.
The thermoelectric power at ambient pressure is as $\sim$3.3 times large as
that at high pressure (at $T \approx$ 80 K),
however the electrical conductivity is reduced due to the larger band gap.
It results in no substantial enhancement in $PF/\tau$ at ambient pressure.

\subsection{Miscibility gap at ambient pressure}

Since FeSb$_{1.97}$As$_{0.03}$ was experimentally
synthesized and was reported to have the much more reduced thermal conductivity
compared to FeSb$_{2}$ \cite{Sun10},
we investigate the miscibility of FeSb$_{2}$
and FeAs$_{2}$ at ambient pressure theoretically.
We take into account the miscibility of FeSb$_{2}$ and FeP$_{2}$ as well.
Considering a mixture of ($1-x$) mole fractions of FeSb$_{2}$
and $x$ mole fractions of FeAs$_{2}$ (FeP$_{2}$)
producing the resultant material FeSb$_{2-2x}$As$_{2x}$ (FeSb$_{2-2x}$P$_{2x}$),
the mixing energy, which is required to obtain the resultant material, is
\begin{eqnarray}
\Delta E_{\text{mix}}(x) &=& E_{f}(\text{FeSb}_{2-2x}\text{X}_{2x})
\nonumber
\\
&-&\Big((1-x) \cdot E_{f}(\text{FeSb}_{2})
+ x \cdot E_{f}(\text{FeX}_{2})\Big),
\nonumber
\\
& &(\text{where X = As or P}),
\end{eqnarray}
where $E_{f}$ are formation energies for the given compounds \cite{formation}.
Together with the material-independent entropy of mixing
$S = -\frac{R}{3} \Big(x\ln(x) + (1-x)\ln(1-x)\Big)$,
where $R$ is the gas constant, the mixing Gibbs free energy is
\begin{eqnarray}
\Delta G_{\text{mix}}(x, T) &=& \Delta E_{\text{mix}}(x)
\nonumber
\\
&+& \frac{RT}{3}\Big(x\ln(x) + (1-x)\ln(1-x)\Big).
\label{Gibbs}
\end{eqnarray}
The boundary of a miscibility gap \cite{Madsen15,Madsen17},
at which the entropy gain compensates the energy cost of mixing,
could be obtained by minimizing Eq.~(\ref{Gibbs}).

Since the energy above hull for FeSbAs (32.2 meV/atom) is much smaller than
that for FeSbP (204.6 meV/atom),
the miscibility gap region of FeSb$_{2-2x}$As$_{2x}$ is quite smaller than
that of FeSb$_{2-2x}$P$_{2x}$.
We would like to note that
the maximal temperature of boundary of miscibility gap
for FeSb$_{2-2x}$As$_{2x}$ and FeSb$_{2-2x}$P$_{2x}$ are
$\sim$1914 and $\sim$5980 K, respectively.
It indicates that the substitution of As for Sb in FeSb$_{2}$ is more favorable
than the substitution of P for Sb in FeSb$_{2}$.

\section{Summary and conclusions}

We investigated the new thermoelectric material FeSbAs,
which is analogous to FeSb$_{2}$ in chemical and structural point of views.
We checked the phase stability of FeSbAs
and found that it can be made thermodynamically stable
at high pressure above $\sim$30 GPa.
Another material analog FeSbP has the same crystal structure as FeSbAs,
however it is more thermodynamically unstable than FeSbAs
and could not be stable even at high pressure.
We also investigated electronic structures of three material analogs
FeSb$_{2}$, FeSbAs, and FeSbP
by using the mBJ method.
Considering the band gap found in FeSb$_{2}$ experimentally,
the mBJ method gives the reasonable electronic structure.
Regarding FeSb$_{2}$ as a member of a family of compounds
(FeSb$_{2}$, FeSbAs, and FeSbP)
we identified that the octahedral rotations could control the size of the
band gap in this series.
We also studied the thermoelectric properties of three material analogs
within our theoretical framework
and found that FeSbAs has the largest Seebeck coefficient among them
above 50 K.
FeSbAs could also have a higher $ZT$ than FeSb$_{2}$
with electron or hole doping.
Hence FeSbAs should be searched experimentally.
More generally, the isovalent substituting Sb with P or As
should be studied.
P is predicted to be much less soluble than As
which indeed has been reported in the literature \cite{Sun10}.

\section{Acknowledgments}
We thank Ran Adler for fruitful discussions.
We also thank Chuk-Hou Yee and Turan Birol for sharing their initial finding
of the crystal structure of FeSbAs at ambient pressure. We are greatful to C. Petrovic,  He
Hua and M. Aronson for discussions and for keeping us informed on their experiments in this
class of materials.  This work was supported by NSF DMREF DMR-1435918.


\section{Appendix: model calculation for transport properties}
In our DFT simulations for temperature dependent transport properties,
the chemical potential was fixed to the middle of the band gap.
However, the chemical potential is a function of temperature.
In semiconducting materials,
the impurity effect also affects the temperature dependence of the chemical potential.
In Appendix, we discuss the temperature behavior of the chemical potential
in the presence of the impurities
and the corresponding thermoelectric power~\cite{eph}.
We also discuss the extremum in the Seebeck coefficient versus temperature curve
observed in experiments for FeSb$_{2}$ at low temperature.

\subsection{Brief review: One-band model with a fixed chemical potential}
First, we briefly review the one-band model with a fixed chemical potential
introduced in Ref.~\cite{Sun13}.
Since the measured thermoelectric power for FeSb$_{2}$ has a negative sign below 50 K,
the dominant charge carrier is electron-type.
For simplicity, the authors of Ref.~\cite{Sun13}
considered a single conduction band model with electron-type carriers.
Furthermore, they assumed a fixed (temperature independent) chemical potential $\mu = -\Delta$
(where $\Delta$ is the activation energy)
and an isotropic parabolic conduction band dispersion
\begin{equation}
\epsilon_{k} = \frac{\hbar^2k^2}{2m^*},
\label{parabolic_dispersion}
\end{equation}
where $m^*$ is the effective mass.
Then the band velocity is obtained by the following relation
\begin{equation}
v_{k}^{\alpha}=\frac{1}{\hbar}\frac{\partial \epsilon_{k}}{\partial k_{\alpha}}
=\frac{\hbar}{m^*}k_{\alpha},
\end{equation}
where $\alpha = x, y, z$.
Transport properties can be computed within the Boltzmann theory
by the following expressions:
\begin{eqnarray}
\sigma_{xx} &=& \frac{2e^2}{V}\displaystyle\sum_{k}\left(-\frac{\partial f}{\partial \epsilon_{k}}\right)v_{k}^{x}v_{k}^{x}\tau_{k},
\\
\alpha_{xx} &=& -\frac{2e}{VT}\displaystyle\sum_{k}\left(-\frac{\partial f}{\partial \epsilon_{k}}\right)v_{k}^{x}v_{k}^{x}(\epsilon_{k}-\mu)\tau_{k},
\end{eqnarray}
where $f$ and $\tau_{k}$ are the Fermi-Dirac distribution function
and the relaxation time, respectively.
The Seebeck coefficient can then easily be calculated
\begin{equation}
S = \frac{\alpha_{xx}}{\sigma_{xx}}.
\end{equation}

Using the fact that the summation of $k$ could be changed into the integral of energy $\epsilon$ with the density of states $D(\epsilon)$, that is, $\displaystyle\sum_{k} \rightarrow \int_{-\infty}^{\infty}d\epsilon D(\epsilon)$, Eqs. (6) and (7) could be rewritten as
\begin{eqnarray}
\sigma_{xx} &=& \frac{4e^2}{3V}\frac{\tau_{0}}{m^*}\int_{-\infty}^{\infty}d\epsilon D(\epsilon)\left(-\frac{\partial f}{\partial \epsilon}\right)\epsilon,
\\
\alpha_{xx} &=& -\frac{4e}{3VT}\frac{\tau_{0}}{m^*}\int_{-\infty}^{\infty}d\epsilon D(\epsilon)\left(-\frac{\partial f}{\partial \epsilon}\right)\epsilon(\epsilon+\Delta),
\end{eqnarray}
where the (direction independent) constant relaxation time approximation
$\tau_{k} \approx \tau_{0}$ is used.
The density of states for the parabolic energy dispersion of Eq. (\ref{parabolic_dispersion}) is
\begin{equation}
D(\epsilon) = V\frac{(2m^*)^{3/2}}{2\pi^2\hbar^3}\epsilon^{1/2}
\quad\quad\quad(\epsilon > 0),
\end{equation}
hence we can estimate Eqs. (9) and (10) for two different limiting cases:
$\Delta \gg k_{B}T$ and $\Delta \ll k_{B}T$.

(i) When $\Delta \gg k_{B}T$,

\begin{eqnarray}
\sigma_{xx} &\simeq& \left(\frac{k_{B}T}{\pi}\right)^{3/2}
\frac{e^2(2m^*)^{1/2}\tau_0}{\hbar^3}\exp\left(\frac{-\Delta}{k_{B}T}\right),
\nonumber
\\
\alpha_{xx} &\simeq& -\frac{e(2m^*)^{1/2}\tau_0}{\pi^{3/2}\hbar^3T}
\left(k_{B}T\right)^{5/2}\exp\left(\frac{-\Delta}{k_{B}T}\right)
\left(\frac{\Delta}{k_{B}T}+\frac{5}{2}\right),
\nonumber
\\
S &\simeq& -\frac{k_{B}}{e}\left(\frac{\Delta}{k_{B}T}+\frac{5}{2}\right).
\end{eqnarray}

(ii) When $\Delta \ll k_{B}T$,

\begin{eqnarray}
\sigma_{xx} &\simeq& \left(\frac{k_{B}T}{\pi}\right)^{3/2}
\frac{e^2(m^*)^{1/2}\tau_0}{\hbar^3}\left(\sqrt{2}-1\right)
\zeta\left(\frac{3}{2}\right),
\nonumber
\\
\alpha_{xx} &\simeq& -\left(\frac{k_{B}T}{\pi}\right)^{3/2}
\frac{e(m^*)^{1/2}\tau_0}{\hbar^3T}
\Big[\left(\sqrt{2}-1\right)\zeta\left(\frac{3}{2}\right)\Delta
\nonumber
\\
& &+\frac{5}{4}\left(2\sqrt{2}-1\right)\zeta\left(\frac{5}{2}\right)k_{B}T\Big],
\nonumber
\\
S &\simeq& -\frac{k_{B}}{e}\left(\frac{\Delta}{k_{B}T}
+\frac{5\left(2\sqrt{2}-1\right)\zeta\left(5/2\right)}
{4\left(\sqrt{2}-1\right)\zeta\left(3/2\right)}\right)
\nonumber
\\
&\simeq& -\frac{k_{B}}{e}\left(\frac{\Delta}{k_{B}T}
+2.833442009\ldots\right),
\end{eqnarray}
where $\zeta(x)$ is the Riemann zeta function.
Note that the Seebeck coefficient $S$ is independent of the relaxation time $\tau_0$
in this approximation.

For both limiting cases,
the Seebeck coefficient is proportional to the inverse of temperature
and does not show the extremum, whereas it does in experiments.

\subsection{Two-band model}

\begin{figure}[t]
\includegraphics[width=8.5 cm]{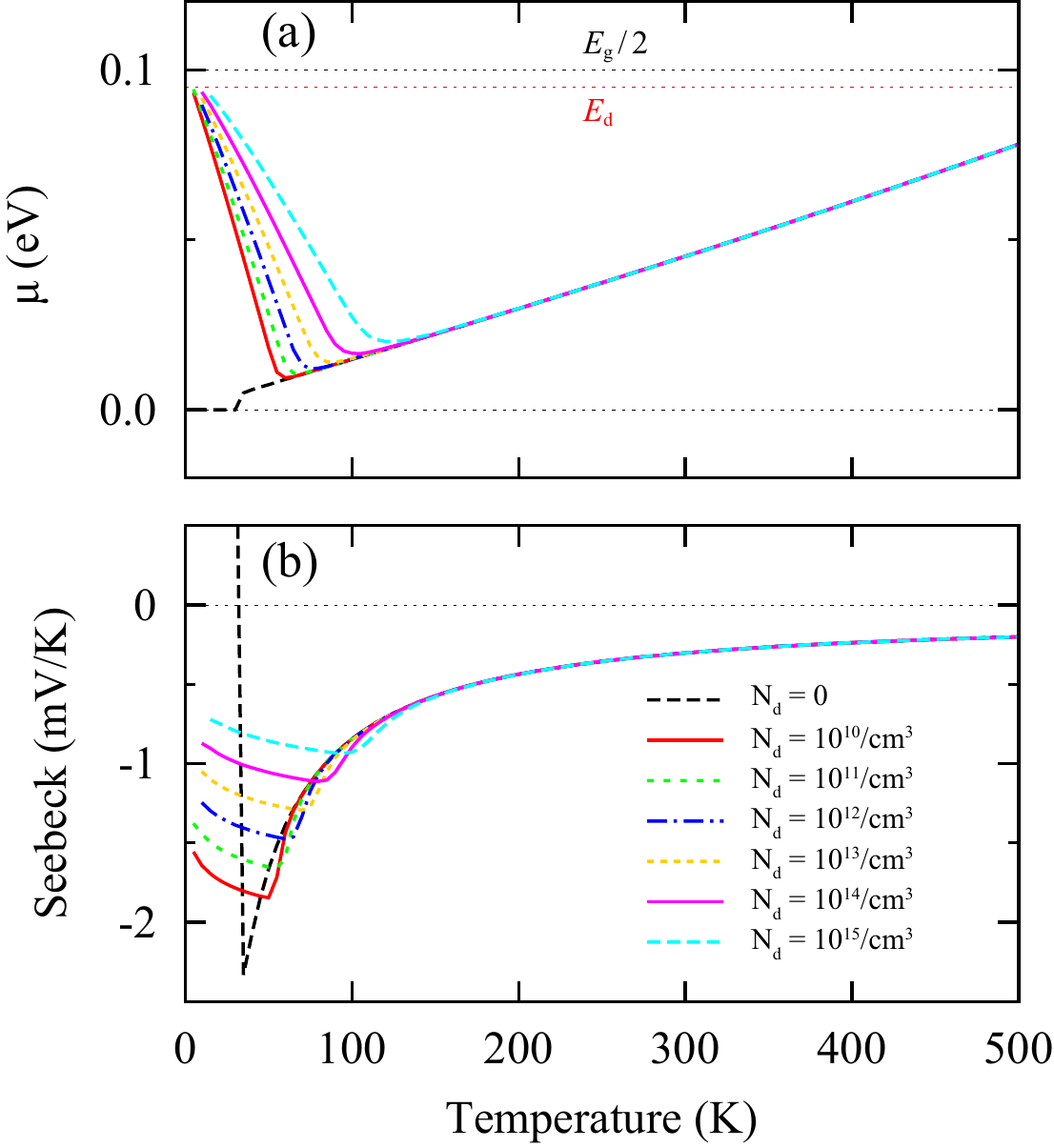}
\caption{(Color Online)
Two-band model for thermoelectric properties with ionized donor impurities.
We set a band gap $E_{g}$ = 0.2 eV,
donor impurity level $E_{d}$ = 95 meV,
valence and conduction band effective masses $m^*_{\text{VB}}$ = 10 $m_{0}$,
$m^*_{\text{CB}}$ = $m_{0}$ (where $m_{0}$ is the electron rest mass),
and cell volume $V$ = 123.673 ${\AA}^3$.
(a) Chemical potential $\mu$ and (b) Seebeck coefficient
as a function of temperature
for different donor impurity concentrations.
}
\label{two-band}
\end{figure}

\begin{figure}[t]
\includegraphics[width=8.5 cm]{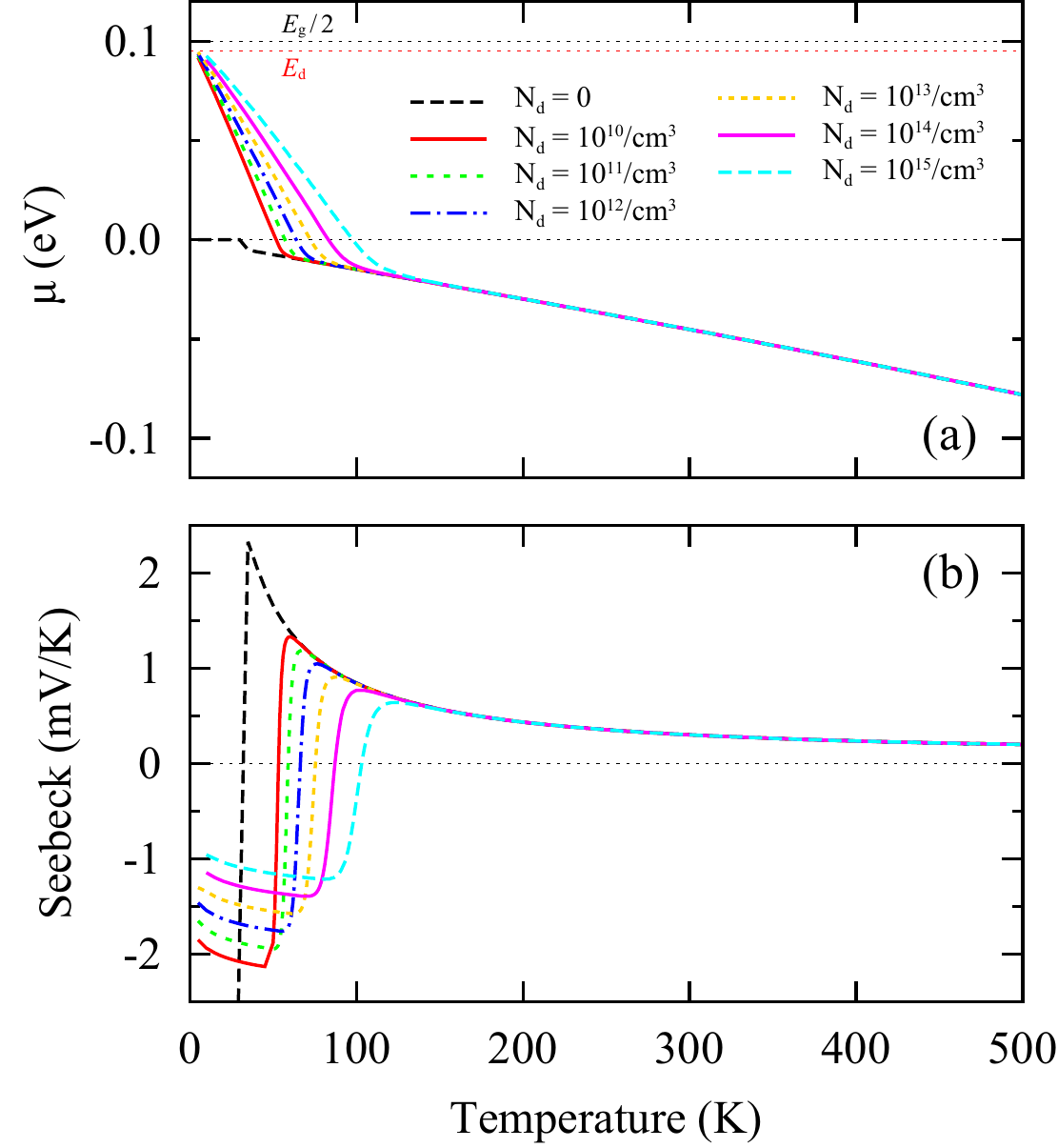}
\caption{(Color Online)
Same as Fig.~\ref{two-band}
except for different valence and conduction band effective masses
in the two-band model.
We set $m^*_{\text{VB}}$ = $m_{0}$ and
$m^*_{\text{CB}}$ = 10 $m_{0}$.
}
\label{two-band2}
\end{figure}

In this subsection, we discuss the two-band model
consisting of one valence and one conduction band
with a band gap of $E_{g}$.
For convenience, we assume simple parabolic band dispersions
for both valence and conduction bands
and positions of the valence band maximum
and the conduction band minimum are $-E_{g}/2$ and $E_{g}/2$,
respectively, in order for the middle point of the band gap to be zero.
Then, the valence and conduction band dispersions are
\begin{eqnarray}
\epsilon_{k}^{\text{VB}} = - E_{g}/2 - \frac{\hbar^2k^2}{2m^*_{\text{VB}}},
\nonumber
\\
\epsilon_{k}^{\text{CB}} = E_{g}/2 + \frac{\hbar^2k^2}{2m^*_{\text{CB}}},
\end{eqnarray}
where $m^*_{\text{VB}}$ and $m^*_{\text{CB}}$ are
the valence and conduction band effective mass, respectively,
and density of states for the valence and conduction bands are
\begin{eqnarray}
D^{\text{VB}}(\epsilon) = V\frac{(2m^*_{\text{VB}})^{3/2}}{2\pi^2\hbar^3}
(-\epsilon-E_{g}/2)^{1/2},
\nonumber
\\
D^{\text{CB}}(\epsilon) = V\frac{(2m^*_{\text{CB}})^{3/2}}{2\pi^2\hbar^3}
(\epsilon-E_{g}/2)^{1/2}.
\end{eqnarray}

To demonstrate that
the Seebeck coefficient of FeSb$_{2}$ has a negative sign
and the extremum at low temperature in experiments,
we allow the temperature dependent chemical potential
and the presence of ionized donor impurities.
(The ionized acceptor impurities make the Seebeck coefficient positive
at low temperature,
which is not the case for FeSb$_{2}$.)
The occupation of ionized donor impurities is
\begin{equation}
N_{d^+} = \frac{N_{d}}{1+2\exp\big(-(E_{d}-\mu)/k_{B}T\big)}
\end{equation}
with donor concentration $N_{d}$ and donor impurity level $E_{d}$.

Occupations of electron ($n$) and hole ($p$) are
\begin{eqnarray}
n = \int_{E_{g}/2}^{\infty}d\epsilon D^{\text{CB}}(\epsilon)f(\epsilon),
\nonumber
\\
p = \int_{-\infty}^{-E_{g}/2}d\epsilon D^{\text{VB}}(\epsilon)\Big(1-f(\epsilon)\Big).
\end{eqnarray}
Then the condition of the charge neutrality $n$ = $p$ + $N_{d^+}$
determines the position of the chemical potential $\mu$,
which is usually as a function of temperature
as shown in Figs.~\ref{two-band}(a) and \ref{two-band2}(a).
In the intrinsic case, $N_{d}$ = 0,
the chemical potential is almost temperature independent
and very close to the middle of the band gap at low temperature.
Above a certain temperature,
the chemical potential shows the linear dependence of temperature
and its slope is determined
by the valence and conduction band effective masses
$m^*_{\text{VB}}$ and $m^*_{\text{CB}}$.
When $m^*_{\text{VB}} > m^*_{\text{CB}}$ (see Fig.~\ref{two-band}(a)),
the linear slope is positive,
hence the chemical potential is close to the conduction band upon heating.
On the other hand, when $m^*_{\text{VB}} < m^*_{\text{CB}}$ (see Fig.~\ref{two-band2}(a)),
the linear slope is negative, that is, the chemical potential is far away
from the conduction band upon heating.

The presence of the ionized donor impurity makes the chemical potential
close to the conduction band at low temperature.
Up to a certain temperature
the chemical potential goes down to the middle of the band gap upon heating,
and then follows the intrinsic high-temperature slope
regardless of the donor impurity concentrations.

Using the obtained temperature dependent chemical potential and Eqs. (6-8),
we can obtain the Seebeck coefficient as a function of temperature
and the result is shown in Figs.~\ref{two-band}(b) and \ref{two-band2}(b).
At finite donor impurity concentrations,
the Seebeck coefficients show an extremum and
the position of the extremum is changed depending on the impurity concentration.
The remarkable difference between Figs.~\ref{two-band}(b) and \ref{two-band2}(b)
is that at finite impurity concentrations
the Seebeck coefficient changes the sign only for $m^*_{\text{VB}} < m^*_{\text{CB}}$
and shows another extremum upon heating~\cite{pdoping}.


\begin{thebibliography}{99}
\bibitem{phonon}
    For phonon calculations, we have used the linear response method \cite{Baroni01}
    implemented in the Quantum Espresso code \cite{QE}.
    PW91 ultrasoft pseudopotentials were used for all atoms.
    We used a 15 $\times$ 15 $\times$ 15 $k$-grid
    and a Hermite-Gaussian smearing of 0.03 Ry for the electronic integration.
    The dynamical matrices were calculated on
    a 2 $\times$ 2 $\times$ 2 phonon-momentum grid.
\bibitem{tau}
    The relaxation time ($\tau$) for FeSb$_{2}$ is obtained
    from the measured $\kappa_{e}$ and calculated $\kappa_{e}/\tau$.
    We assume that relaxation times for FeSb$_{2}$, FeSbAs and FeSbP
    are almost identical.
\bibitem{formation}
    The formation energies for FeSb$_{2}$, FeSbAs (FeSbP), and FeAs$_{2}$ (FeP$_{2}$)
    are calculated by the VASP GGA(PBE) functional
    and then those for intermediate concentrations of FeSb$_{2-2x}$As$_{2x}$
    (FeSb$_{2-2x}$P$_{2x}$) are obtained by interpolation.
\bibitem{eph}
    We do not consider the electron-phonon coupling
    that could affect the transport properties.
    Note that the sizable electron-phonon coupling is measured from
    polarized Raman scattering spectra of the Fe$_{1-x}$Co$_{x}$Sb$_{2}$
    and Fe$_{1-x}$Cr$_{x}$Sb$_{2}$ single crystals~\cite{Lazarevic10}.
\bibitem{pdoping}
    In case of the presence of the ionized acceptor impurity ($p$-type doping),
    the opposite behavior is expected.
    When $m^*_{\text{VB}} > m^*_{\text{CB}}$, the Seebeck coefficient changes the sign from a positive to a negative value upon heating.
    When $m^*_{\text{VB}} < m^*_{\text{CB}}$, the Seebeck coefficient always has a positive value for a finite impurity concentration.
\end{thebibliography}
\end{document}